\documentclass[10pt,pra,twocolumn,showpacs,floatfix,superscriptaddress]{revtex4-1}
\usepackage{graphicx}
\usepackage{epstopdf}
\usepackage[colorlinks]{hyperref}
\hypersetup{
 colorlinks=true,
 citecolor=blue,
 linkcolor=black,
 urlcolor=blue}
\usepackage{amssymb}
\usepackage{times}
\usepackage{amsmath,mathrsfs}
\usepackage{amsthm}
\usepackage{dsfont}
\usepackage{amsfonts}%

\newcommand{\ket}[1]{|#1\rangle}

\renewcommand{\sec}[1]{\hyperref[sec:#1]{Sec.~\ref{sec:#1}}}
\newcommand{\eq}[1]{Eq. (\ref{eq:#1})}
\newcommand{\fig}[1]{\hyperref[fig:#1]{Fig.~\ref{fig:#1}}}
\newcommand{\dx}{dx\,}
\newcommand{\id}{\mathds 1}
\newcommand{\etal}{\mbox{\emph{et al.\ }}}

\begin{document}
\title{Repeat-until-success cubic phase gate for universal continuous-variable quantum computation}
\author{Kevin Marshall}
\affiliation{Department of Physics, University of Toronto, Toronto, M5S 1A7, Canada}
\author{Raphael Pooser}
\affiliation{Quantum Information Science Group, Oak Ridge National Laboratory, Oak Ridge, Tennessee 37831, U.S.A}
\affiliation{Department of Physics and Astronomy, The University of Tennessee, Knoxville, Tennessee 37996-1200, U.S.A.}
\author{George Siopsis}
\affiliation{Department of Physics and Astronomy, The University of Tennessee, Knoxville, Tennessee 37996-1200, U.S.A.}
\author{Christian Weedbrook}
\affiliation{QKD Corp., 60 St.~George St., Toronto, M5S 1A7, Canada}
\date{\today}
\begin{abstract}
In order to achieve universal quantum computation using continuous variables, one needs to jump out of the set of Gaussian operations and have a non-Gaussian element, such as the cubic phase gate. However, such a gate is currently very difficult to implement in practice. Here we introduce an experimentally viable `repeat-until-success' approach to generating the cubic phase gate, which is achieved using sequential photon subtractions and Gaussian operations. We find that our scheme offers benefits in terms of the expected time until success, although we require a primitive quantum memory.
\end{abstract}
\maketitle

\section{Introduction}
Modern digital computers operate by manipulating discrete information, ones and zeros, and are able to perform complicated tasks by first breaking them down into more elementary operations.  The first notions of quantum computation proceed in a largely analogous manner where the elementary operators can be classified into two groups \cite{nielsen}: Clifford and non-Clifford gates acting on discrete states.  In order to achieve universal quantum computation one needs elements from both of these sets \cite{gottesman}; the set of Clifford operations can typically be implemented in a laboratory with relative ease, while the non-Clifford gates are the main obstacle in achieving universal quantum computation.  One can also consider quantum computation over continuous-variables (CVs) \cite{Braunstein05,weedbrook} to obtain the same speedup, in a manner similar to classical analogue computation, and in fact, this framework may have inherent benefits \cite{Lloyd99}.  However, CV systems also have a set of difficult to implement operations that are required for universal quantum computation, namely non-Gaussian gates.  In addition, there exist hybrid approaches which seek to exploit the advantages of both discrete and CV systems to offer the best of both worlds \cite{Lloyd03,Loock11,Furusawa11}.  These reasons serve as our motivation for understanding how to implement non-Gaussian gates in a practical manner.

In addition to the readily available set of Gaussian operations, corresponding to Hamiltonians of first and second powers of the quadrature operators $\hat x$ and $\hat p$, we only require access to a Hamiltonian of at least third power in the quadrature operators to allow us to approximate a Hamiltonian that is an arbitrary polynomial of the quadrature operators \cite{Lloyd99}.  In particular, we consider an elementary operation known as the cubic phase gate, given by $U(\gamma)=e^{i\gamma \hat x^3}$.  Presently, one can directly implement this Hamiltonian only with $\gamma\ll 1$ due to very low interaction strengths, and the noise one introduces by doing so makes this unsuitable for quantum computation \cite{Bartlett}.  Alternative approaches seek to use non-linearity from photon subtraction or addition, or non-Gaussian measurements such as photon counting \cite{GKP,Marek}.

In this paper, we propose a method capable of approximating the cubic phase gate to arbitrary accuracy while only requiring sequential photon subtractions and Gaussian operations.  Our method is inherently probabilistic, but can be operated in a repeat-until-success fashion using technology that is available today, and we show that the time until success of our method scales only as slow as $1/p$ where $p$ is the probability of a successful photon subtraction.  Furthermore, we require only standard photon detectors which can distinguish between vaccuum and the presence of one or more photons, such as an avalanche photodiode (APD), and do not need homodyne detectors.  Previous experiments in the field of continuous-variables have demonstrated the implementation of a variety of key elements such as: entangling gates \cite{yokoyama14}, dynamic squeezing \cite{Miyata}, scalability \cite{yokoyama13,pysher,chen14} and state preparation \cite{takeda}, however the cubic phase gate remains a key obstacle.

This paper is organized as follows.  In \sec{background} we motivate the necessity of the cubic phase gate by discussing its role in universal quantum computation and we show a method of approximating the gate with a sequence of non-deterministic operations.  Next, we discuss an ideal and realistic method of implementing our decomposition in \sec{implementation}.  In \sec{realistic} we study the effect of detector imperfections as well as choosing a finite truncation in our approximation and compare these results to the ideal cubic phase gate.  An explicit experimental implementation including a discussion of the repeat-until-success nature of the photon subtraction is given in \sec{experimental}.  In \sec{compare} we compare our proposal to two other implementations of the cubic phase gate, and finally we provide concluding remarks in \sec{conclusion}.

\section{Background}
\label{sec:background}
The cubic phase gate \cite{GKP} is an essential component in CV universal quantum computation as it allows us to approximate arbitrary Hamiltonians \cite{Lloyd99}.  This comes from the fact that if one is able to implement operators $\hat A$ and $\hat B$, then one can approximate the operator $i[\hat A,\hat B]$ using the relation \cite{Lloyd99}
\begin{align}
\label{eq:commutator}
e^{i\hat A t}e^{i\hat B t}e^{-i\hat A t}e^{-i\hat B t}\approx e^{-[\hat A,\hat B]t^2}+\mathcal O(t^3).
\end{align}
Specifically this allows us to construct Hamiltonians consisting of monomials of $\hat x$ of degree higher than three as \cite{Sefi}
\begin{align}
\hat x^m= \frac{-2}{3(m-1)}\left[\hat x^{m-1},\left[\hat x^3,\hat p^2\right]\right],
\end{align}
from which we can construct arbitrary polynomials using the relationship
\begin{align}
\hat x^m\hat p^n+\hat p^n \hat x^m&=\frac{-4i}{(n+1)(m+1)}\left[\hat x^{m+1},\hat p^{n+1}\right]\nonumber\\
&-\frac{1}{n+1}\sum_{k=1}^{n-1}\left[\hat p^{n-k},\left[\hat x^m,\hat p^k\right]\right].
\end{align}

In order to approximate the cubic phase gate we first consider a decomposition
\begin{align}
U_N(\gamma)=\left(1+i\frac{\gamma}{N}\hat x^3\right)^N,
\end{align}
clearly as $N$ becomes large this well approximates the cubic phase gate to within terms of the order $\mathcal O(1/N)$.  We further decompose this operation as
\begin{align}
1+i\frac{\gamma}{N}\hat x^3&=\mathcal U_0\mathcal U_1\mathcal U_2,
\end{align}
where $\mathcal U_l=1+\gamma_l\hat x$ and $\gamma_l=\exp[i\pi(4l+1)/6](\gamma/N)^{1/3}$.  Note that, \begin{align}
\left|\mathcal U_0\mathcal U_1\mathcal U_2\right|^2&=1+\frac{\gamma^2}{N^2}\hat x^6,
\end{align}
so that for $N$ sufficiently large, compared to values of $x$ where $\psi(x)$ has non-negligible support, this operation, when successful, will approximately preserve the norm of our state.  Our goal is to now find a method of implementing $\mathcal U_l$, as we can approximate the cubic phase gate to within error $\mathcal O(1/N)$ by $N$ applications of these three operators.
\section{Implementation}
\label{sec:implementation}
We would like to implement the cubic phase gate on an arbitrary state $\ket\psi=\int \dx \psi(x)\ket x$ and to do so we have noted that it is enough to find a method of implementing the various $\mathcal U_l$ operators. To do so we first prepare a coherent state $\ket{\alpha_1}_R=D(\alpha_1)\ket 0_R$, where $D(\alpha_1)=\exp(\alpha \hat a^\dagger_R-\alpha^*\hat a_R)$ is the familiar displacement operator and where the subscript $R$ denotes a resource mode.  We then interact our state of interest with the coherent state under the two-mode operator $U_2(\beta)=\exp[(\beta\hat a^\dagger_R-\beta^*\hat a_R)\hat x]$; this gate is known as the quantum non-demolition (QND)-gate and its implementation requires two offline squeezed ancilla states \cite{Yoshikawa}.  After this interaction we arrive at the state
\begin{align}
\ket{\Psi}&=\int \dx \psi(x) U_2(\beta_1)D_R(\alpha_1)\ket x\ket 0_R\nonumber\\
&=\int\dx\psi(x)\ket x\ket{\alpha_1+\beta_1 x}_R\nonumber\\
\label{eq:Psi}
&=\int\dx\psi(x)\ket x\ket{\alpha_1(1+\gamma_l x)}_R,
\end{align}
where we have chosen $\beta_1=\gamma_l\alpha_1$ and where $\alpha_1\in\mathbb R$ is an unspecified and tunable parameter.  We proceed by describing two different methods of obtaining the desired result; one method is theoretically simple but experimentally challenging, while the other can feasibly be demonstrated with currently available technology.
\subsection{Ideal implementation}
First we will analyze the ideal implementation to convey the main concept in our approach.  Given the state in \eq{Psi} we apply a non-demolition measurement on the resource mode and condition on the outcome where we project onto the space orthogonal to $\ket 0$, i.e., the outcome where $P_{\bar 0}=\hat \id - |0\rangle\langle 0 |$ is the corresponding projector.  After this projection we are left with the state
\begin{align}
P_{\bar 0}\ket{\alpha_1(1+\gamma_l x)}_R&=\sum_{k=1}^\infty \frac{1}{\sqrt{k!}}\left[\alpha_1(1+\gamma_l x)\right]^{k}\ket k_R\\
&\approx \alpha_1(1+\gamma_l x)\ket 1_R
\end{align}
where $\alpha_1\in \mathbb R$ is some suitably chosen constant given that the error in this approximation is negligible for small $x$ and significant for large values of $x$; in practice we desire some \emph{a priori} information about $\psi(x)$.  With this transformation, the final state is of the form
\begin{align}
\ket{\Psi'}&\propto\int \dx \psi(x) (1+\gamma_l\hat x)\ket x \ket 1_R\nonumber\\
&\propto\int\dx \psi(x)\mathcal U_l\ket x\ket 1_R,
\end{align}
which is the desired outcome.  This approach is probabilistic in that the non-demolition measurement may fail, however if it does fail one can simply discard the mode and attempt the procedure again.
\subsection{Realistic implementation}\label{sec:realistic}
The non-demolition measurement in the previous section is experimentally a very challenging task, however it serves as a solid motivation towards finding a more practical approach.  In this section we discuss a more feasible approach based on sequential photon subtraction, namely we take the state in \eq{Psi} and pass the resource mode through a beam splitter of high transmittance $T$ mixing it with a vacuum input $\ket 0_b$ to obtain
\begin{align}
\ket{\Psi}\ket 0_b\rightarrow \int \dx \psi(x) \ket x \ket{\sqrt T \zeta_l}_R \ket{-\sqrt{1-T}\zeta_l}_b,
\end{align}
where $\zeta_l=\alpha_1(1+\gamma_l x)$.  We require the ancillary mode to be weak so that the $b$-mode is well approximated by $\ket{-\sqrt{1-T}\zeta_l}_b\approx \ket 0_b-\sqrt{1-T}\zeta_l\ket 1_b$.  Next we send the $b$-mode to a photodetector which acts, in this two-dimensional subspace, as the projector $|1\rangle\langle 1 |$ if it clicks, and otherwise as $|0\rangle\langle 0 |$.  If there is no click then we are left with the state $\int \dx \psi(x) \ket{x}\ket{\sqrt T \zeta_l}_R\ket 0_b$ and the $b$-mode decouples so that we can reattempt the last step; in this way we are able to operate in a repeat-until-success fashion without a change in the state except for an attenuation $\sqrt T\approx 1$.  If the probability of subtracting a single photon is given by $p$, which will depend on the parameters and state of interest, then this scheme will succeed after $\sim 1/p$ trials, and thus we can approximate the cubic phase gate to within terms of order $\mathcal O(1/N)$ within a running time that scales as $\sim 3N/p$ .  Suppose it takes $M$ attempts to remove one photon from the resource mode so that the new state of the system is
\begin{align}
\label{eq:steady}
\ket{\Psi'}\propto\int\dx\psi(x) (1+\gamma_l \hat x)\ket x \ket{T^{M/2}\alpha_1(1+\gamma_l x)}_R\ket 1_b.
\end{align}
We can decouple the resource mode by applying another QND gate $U_2(-T^{M/2}\alpha_1\gamma_l)$ to obtain
\begin{align}
\ket{\Psi''}\propto\int\dx\psi(x)\mathcal U_l\ket x\ket{T^{M/2}\alpha_1}_R\ket{1}_b,
\end{align}
which is the desired gate after we trace out the last two modes which are only part of a product state with our mode of interest.

\begin{figure}[htp]
\centering
\includegraphics[scale=0.8]{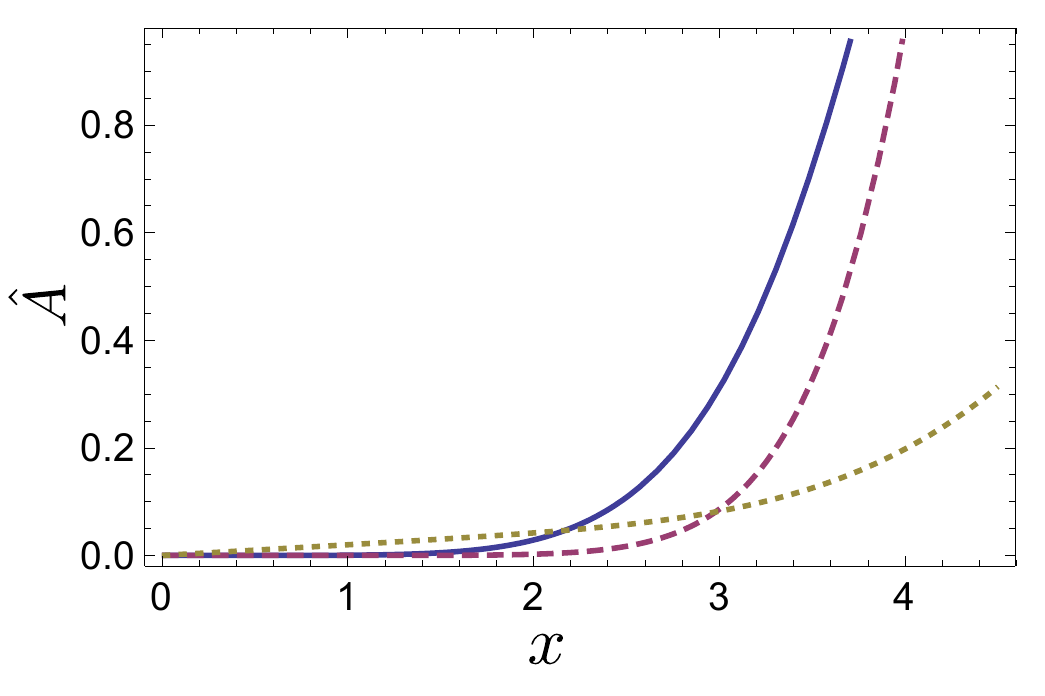}
\caption{(Color online) The effect of imperfect detectors is shown where the operator $\hat A=U_N(\gamma)-e^{i\gamma\hat x^3}$ is the difference between the approximation and ideal cubic phase operators.  The expectation value, both real (solid) and imaginary (dashed) parts, as well as the standard deviation (dotted) of this operator, in a position eigenstate $\ket x$, is plotted for $\gamma=0.03$ \cite{Marek,Mitsuyoshi} and $N=1$.  We consider a detector with an efficiency of $90\%$, a dark count rate of $100$Hz and a timing resolution of $100$ps; such a detector is within the reach of current technology \cite{marsili2013,rosenberg2013,miki2013}.  We see that for small values of $x$ the effect of detector imperfections on our approximate cubic phase gate are small and we still approximate the ideal gate closely.}\label{fig:approx}
\end{figure}

\begin{figure}[htp]
\centering
\includegraphics[scale=0.8]{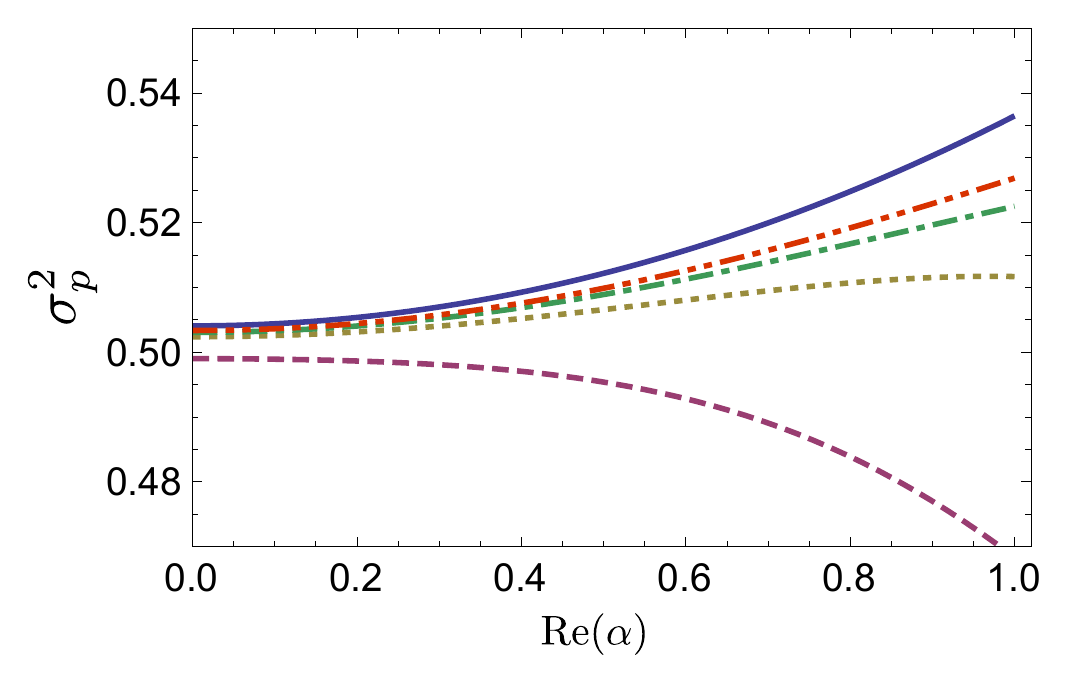}
\caption{(Color online) The variance $\sigma_p^2$ is plotted for the ideal (solid) cubic phase gate as well as $U_N(\gamma)$ for $N=1$ (dashed), 3 (dotted), 5 (dash-dot), 7 (dash-double dot); the strength of the gate is given by $\gamma=0.03$.  The input state is chosen to be a coherent state with constant imaginary part Im$(\alpha)=0.25$ where the real part varies with the horizontal axis.  Additional squeezing, omitted in this plot, can be utilized to improve the strength or quality of the gate; here we are interested only in the differences as one varies $N$ and we see that as $N$ increases we obtain a closer approximation.  Deviation from the ideal curve will result in extra variance in the momentum quadrature, compared to the ideal curve, after one squeezes the mode.}\label{fig:varp}
\end{figure}

This approach assumes an ideal photon detector that is capable of only distinguishing between vacuum and the presence of one or more photons.  We can model imperfections in such a detection scheme by considering the POVM $\{\hat \Pi_0,\hat \id -\hat \Pi_0\}$ where
\begin{align}
\hat \Pi_0&=\sum_{m=0}^\infty e^{-\nu}(1-\eta)^m|m\rangle\langle m|,
\end{align}
and where $\eta$ is the detection efficiency, while $\nu$ is the rate of dark counts \cite{Barnett}.  For our purposes, an imperfect detection efficiency $\eta< 1$ will result in cases where we have successfully subtracted a photon but we are unaware of this fact.  This means that we will obtain the desired transformation in \eq{steady}, however, since we do not recognize it, we will proceed in attempting further photon subtractions; this will effectively increase the power on the factor $(1+\gamma\hat x)^{k>1}$ in an undesired manner.  Alternatively, for $\nu\neq 0$ there will exist false positives where we believe we have subtracted a photon when we have not.  In this case, the protocol will proceed where one of the $\mathcal U_l$ operators is replaced by the identity operation; the effect of these errors is plotted in \fig{approx}.
We can also consider consider the effect of the cubic phase gate on the first two moments of $\hat x,\hat p$.  We find good agreement for the first moments between the ideal gate and our approximation.  The variance of momentum is plotted in \fig{varp} to demonstrate that for increasing values of $N$ one obtains a result closer to the ideal operation.  One can also increase the strength of the approximate cubic phase gate by squeezing, for example when $N=1$ one finds that an effective strength $\gamma_{eff}\sim 0.1$ can be achieved with $\gamma=0.03$ without greatly degrading the quality of the gate \cite{Marek,Mitsuyoshi}.
\section{Experimental setup}\label{sec:experimental}
Section \ref{sec:implementation} discussed the theoretical implementation of the cubic phase gate by outlining the full operator for a single factor in $\mathcal U_0\mathcal U_1\mathcal U_2$. The full gate can then be realized by applying the three operators sequentially to the target state, resulting in an additional power in the phase for each factor. In this section, we outline experimentally how to achieve a single operator for a linear phase gate, which can then be applied experimentally in sequence for a cubic, or higher polynomial, phase gate.

\begin{figure*}[ht]
\includegraphics[width=7in]{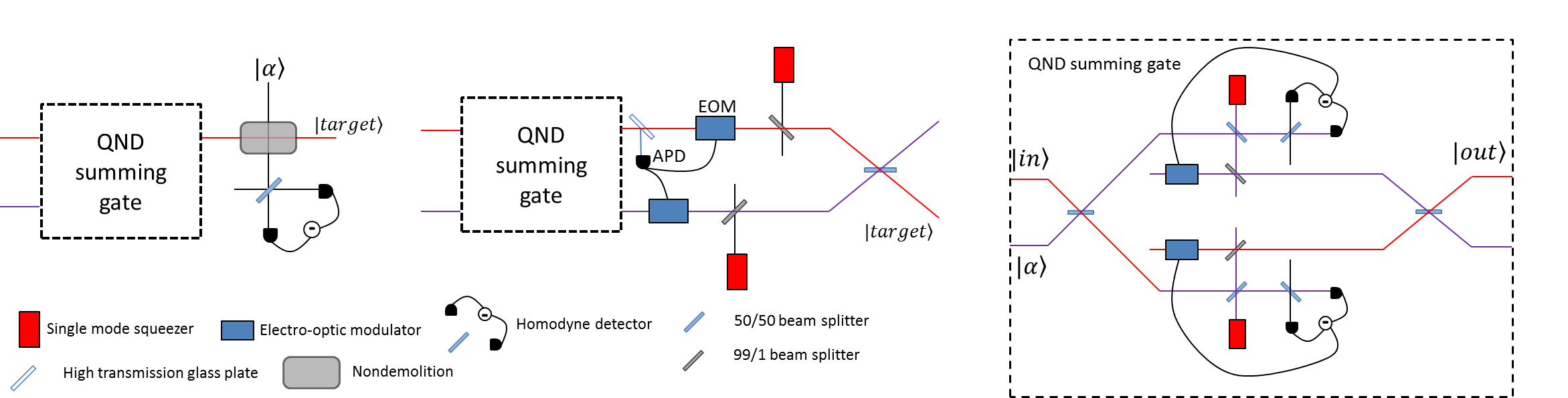}
\caption{(Color online) Experimental implementation of a single $\mathcal U_l$ operation in the cubic phase gate decomposition. Left: after application of the two-mode unitary a QND measurement is made to project onto the subspace orthogonal to $|0\rangle\langle 0|$. This could consist of an inverse V-STIRAP scheme via strong atomic coupling to a photonic cavity mode (cavity QED)\cite{jeffers_ND}, or via cross-phase modulation. However, in the case of cross-phase modulation a phase shift can be registered, but not its magnitude, in order to avoid projection onto a specific number state. Middle: After the QND measurement, the resource state is sent through a weak beam splitter and an avalanche photodiode registers a click to signal a photon-subtraction measurement. The QND summing gate is then run in reverse, where the photon detection triggers inverse operations on the phase modulators and squeezers compared to the original QND summing gate. If the squeezers in the inverse gate are kept phase locked to the first set in the previous measurement, homodyne detectors are unecessary here as the APD serves as the conditional measurement and the squeezing angles are known, allowing for deterministic feed forward control. Right: the QND summing gate consisting of feed-forward phase modulation and offline squeezing resources. }\label{fig:setup}
\end{figure*} 

\subsection{Quantum non-demolition measurement}
Two distinct non-demolition measurements are applied in the ideal gate implementation. First, the two-mode unitary  $U_2(\beta) =  \text{exp}[(\beta \hat{a}_R^\dagger - \beta^\ast \hat{a}_R) \hat{x}] $ is implemented via the non-demolition phase gate \cite{furusawa_QND,Andersen_QND} (see \fig{setup}).
A subsequent non-demolition measurement must be performed in order to project the state onto the subspace complementary to $|0\rangle\langle 0|$ without projecting into a specific number eigenstate. Such a measurement has been theoretically proposed using the inverse V-STIRAP scheme for single photon emission \cite{jeffers_ND}. Briefly, the cavity-QED scheme used in V-STIRAP is inverted: an atom in a specific ground state sublevel interacts with a cavity field containing an unknown number of photons. The atom and cavity modes are entangled; allowing a measurement of which ground state sublevel the atom occupies to determine the photon field state. One sublevel denotes the vacuum state for the cavity field, while the other sublevel denotes the original cavity field with a single photon subtracted. This scheme requires our resource state to serve as the cavity mode in a cavity-QED scheme. This is experimentally challenging but should be achievable with current technology. Alternatively, one may perform a variation of cross-phase modulation. A precise measurement of a phase shift on a reference beam would in fact project the resource state onto a specific photon number state, rather than onto the complement of $|0\rangle\langle 0|$. However, if the presence of a phase shift can be determined, but not its magnitude, one could approximate the inverse V-STIRAP measurement.

\subsection{Photon subtraction measurement}
Alternatively, one may approximate the QND projection via single photon subtraction and post-selection (see \fig{setup}). One may try the photon subtraction as many times as necessary, such that the success probability of the gate is directly related to the beam splitter transmission and avalanche diode efficiency. However, the nonunitary gate enacted by photon subtraction requires only a single photon be removed, resulting in a trade-off between wait time and fidelity. We note that the time dependence has been removed in all the operators presented in the manuscript thus far. We are therefore describing the steady-state super-operator, corresponding to working in an interaction picture where we have removed the free evolution $H_0=\hbar\omega \hat a^\dagger\hat a$, and therefore the steady-state form of $|\Psi\rangle$ in \eq{steady}. Upon a successful photon subtraction, the super-operator jumps to an operator that contains the nonunitary gate. The photon subtraction signal serves as the feed-forward mechanism for the next stage of the gate.

\begin{figure}
\includegraphics[width=3.5in]{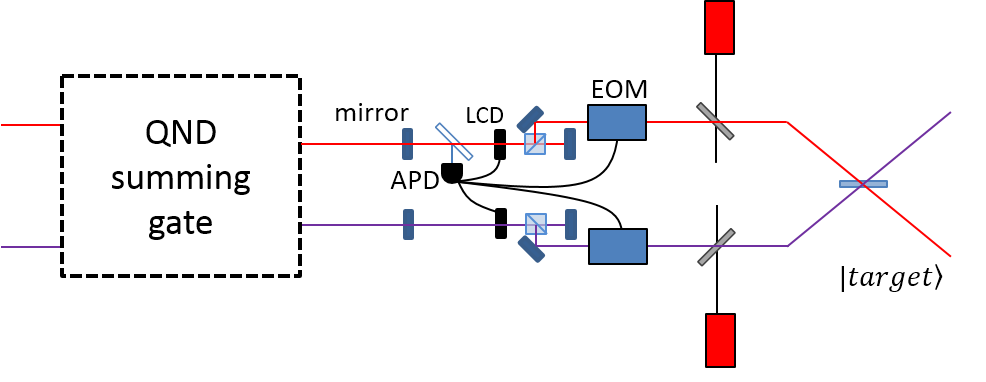}
\caption{(Color online) Repeat-until-success scheme in which a pulsed resource state remains in a cavity until a photon detection is registered.}\label{fig:circ}
\end{figure}

With the initial QND summing gate running in steady state, the resource state is deterministically generated before the subtraction operation. Therefore, one waits until a photon has been subtracted and a new steady state is achieved before feed-forward operations in the inverse QND gate. This scheme is remarkably simple for CW beams: one has the same resource state at all points in space-time of the experimental setup, and the state only changes when the APD registers a click and the subsequent feed-forward operations are performed. For a pulsed light source, the repeat-until-success scheme would consist of coupling into a circulator where the APD is located, followed by a feed-forward signal to couple out of the circulator (see \fig{circ}). The circulator can be implemented in optical fiber or in a polarization sensitive ring cavity. A liquid crystal wave retarder serves as a switch to release the resource state once a successful photon subtraction is registered.

The final step is to invert the QND summing gate. By phase-locking all of the offline squeezed light resources in the experiment, the second homodyne detection can be eliminated, as the first homodyne detection will also feed-forward onto the squeezing angle of the inverse gate's single-mode squeezers. The single photon subtraction gate serves as a switch to perform inverse phase modulation and inverse squeezing operations from the previous QND gate.

After this final step, the resource state is left in a phase-dependent state with a single power in phase. The experimental setup outlined above can be performed three times in sequence in order to realize a cubic phase gate. Finally, the task remains to utilize this gate during a universal quantum computation in order to achieve the required non-Gaussian operation. It is sufficient to teleport the gate onto one of the modes in the resource cluster state in a one-way CV QC scheme, for example \cite{gu09}.

\section{Comparison to other schemes}\label{sec:compare}
In this section we explore the relationship between our proposal and alternative realizations of the cubic phase gate.  Direct implementation of the cubic phase gate would require the use of third order or higher optical nonlinearities.  However, these higher order nonlinearities are so weak that in practice it is difficult to use them to implement quantum gates due to the comparatively large amount of noise and loss.  Fortunately, it was shown by Gottesman, Kitaev, and Preskill (GKP) that one can take advantage of the effective nonlinearity present in a photon counting measurement \cite{GKP}.  In the GKP-scheme, one implements the cubic phase gate by first constructing the, ideal and not normalizable, cubic phase state \cite{GKP} $\ket \xi=\int\dx e^{i\xi x}\ket x$.  Given this resource state, one is able to implement the cubic phase gate using only additional Gaussian operations and homodyne measurement; the main difficulty lies in the creation of the resource state.  

To create a cubic phase state we first prepare an imperfect EPR pair which we take to be Gaussian
\begin{align}
\ket{\psi_{\sigma_x,\sigma_p}}=&\left(\frac{\sigma_p}{\pi\sigma_x}\right)^{-\frac{1}{2}}\int dx_1 dx_2\,\exp\left[-\frac{1}{2}\sigma_p^2\left(\frac{x_1+x_2}{2}\right)^2\right]\nonumber\\
&\times\exp\left[-\frac{1}{2}\left(x_1-x_2\right)^2/\sigma_x^2\right]\ket{x_1,x_2},
\end{align}
with $\sigma_p,\sigma_x\ll 1$.  We then mix the second oscillator with a coherent beam of light to obtain a large shift in momentum $\ket\psi\rightarrow e^{i\alpha \hat q}\ket \psi$, $\alpha\gg \sigma_p^{-1},\sigma_x^{-1}$, before we make a measurement of the number of photons.  After detecting $n$ photons and making some simplifying approximations, namely using the WKB approximation for the eigenstates of the number operator, we find that the state of the first mode is given by
\begin{align}
\psi^{(n)}_1(x_1)\propto \exp\left[i\frac{x_1^3}{6\sqrt{2E}}-i\left(\sqrt{2E}-\alpha\right)x_1+\mathcal O\left(\frac{x_1}{\alpha}\right)\right]\nonumber,
\end{align}
where $E=n+1/2$ and we are most likely to get get a result in the range $n+1/2\sim 1/2(\alpha\pm\sigma_x^{-1})^2+1/2\sigma_p^{-2}$.  This is a good approximation to the cubic phase state provided that $\alpha$ is large enough since we can eliminate the linear term with a Gaussian operation.  Although the  coefficient $\gamma'$ of the cubic phase state in the above approximation is of the order $n^{-1/2}$ we wish to be able to implement $U(\gamma)$ where the coefficient is of the order one.  To do so GKP showed that it is enough to be able to squeeze the phase state with a squeezing parameter $r=\gamma/\gamma'$.  The fact that the ideal cubic phase state is not a physical state, and one can only make approximations of it, has been analyzed where it has been shown that one requires squeezing beyond what is experimentally possible with current technology, although this limitation might be overcome by using other methods of generating finite superpositions of Fock states \cite{Ghose}.

Another alternative to counting photons, a non-Gaussian measurement, is to appeal to experimentally feasible non-Gaussian operations such as photon subtraction or addition.  This is the approach taken by Marek, Filip, and Furusawa \cite{Marek}, who employ the principles of GKP's approach where one indeed overcomes the squeezing limitation by instead using a resource state that is finite in the number basis.  This approach proceeds by creating a squeezed state $S(r)\ket 0_R=(\pi r)^{-1/4}\int \dx\exp(-x^2/2r)\ket x$ where the state approaches the ideal form as $r\rightarrow \infty$.  The cubic phase gate $U(\gamma)\approx 1+i\gamma \hat x^3-\gamma^3\hat x^6/2$ can be approximated by its Taylor expansion, and we note that the lowest order expansion which allows the commutation tricks in \sec{background} requires us to keep terms up to $\hat x^6$; this would require six photon subtractions.  In the Marek \etal proposal, the authors look specifically at the lowest order approximation $\mathcal O_3= 1+i\gamma\hat x^3$ where one then applies the operator $1+i\gamma\hat x_R^3$
to the resource state to obtain
\begin{align}
&\left(1+i\gamma\hat x^3_R\right)S(r)\ket 0_R\nonumber\\
&=S(r)\left[\ket 0_R+\gamma'\frac{3}{2\sqrt 2}\ket 1_R+\gamma'\frac{\sqrt 3}{2}\ket 3_R\right],
\end{align}
where $\gamma'=\gamma r^{-3/2}$.  This state can be generated by the proper sequence of three photon subtractions and displacements.  Given that we have this resource state at our disposal we apply the QND gate $U_2'=e^{i\hat p\hat x_R}$ to obtain the state
\begin{align}
\ket{\Psi}=\int dxdx_R\, \psi(x) e^{-\frac{(x+x_R)^2}{2r}}\left[1+i\gamma (x+x_R)^3\right]\ket x\ket x_R.
\end{align}
If we perform a homodyne measurement of $\hat x_R$ and obtain the outcome $q$ the state of the first mode will collapse to
\begin{align}
\psi'(x)=e^{-\frac{(x+q)^2}{2r}}\left[1+i\gamma(x+q)^3\right]\psi(x),
\end{align}
and as $r\rightarrow \infty$ the exponential term will vanish leaving us with only the cubic term.  If $q=0$ this is the desired outcome and we are finished, however if this is not the case we can perform the feed-forward unitary $U_{FF}=\exp[-i\gamma q^3-3i\gamma(\hat x+q)\hat x q]$ to arrive at the state $\psi'(x)=[1+i\gamma x^3 +\mathcal O(1)]\psi(x)$; this approximation is only valid for sufficiently small $x,q$.  This approach can be generalized to work with higher order approximations $\mathcal O_n$ of the cubic phase gate that simply implement more terms in the Taylor expansion.

Our approach is similar to the Marek \etal approach where one further decomposes the $(1+\gamma\hat x^3)$ operation into a sequence of three non-deterministic operations; for the $N=1$ case both approaches result in the same approximation $U_N(\gamma)$. This idea is noted by the authors in Ref. \cite{Marek}, where they point out that they are unable to make use of this trick since they require deterministic feed-forward in their approach.  However, our protocol requires only a photon detector which can distinguish between vaccuum and one or more photons, and one can operate in a repeat-until-success fashion.   An overview of the requirements and a comparison of the various schemes is presented in \hyperref[tab:comparison]{Table.~\ref{tab:comparison}}.
\begin{table*}[tb]
\centering
\begin{tabular}{|p{2.5cm}|p{4cm}|p{4cm}|p{4cm}|}\hline
&\multicolumn{1}{|c|}{GKP} &\multicolumn{1}{|c|}{Marek \etal} &\multicolumn{1}{|c|}{Ours} \\\hline
Resource & Two-mode squeezed vacuum & Single-mode squeezed vacuum & Coherent state\\
Detectors & Photon number resolving & Homodyne+photon detector & Photon detector\\
Deterministic & Yes, with offline resource & Yes, with offline resource & Repeat-until-success (state not destroyed)\\
Running Time & N/A & $\sim1/p^3$ & $\sim 1/p$\\
Major Obstacle & High squeezing & Simultaneous photon subtraction/resource state engineering & Decoherence and extra noise while running the subtraction loop\\
\hline
\end{tabular}
\caption{Comparison of three cubic phase gate implementations.  Note that the Marek \etal approach is deterministic if one can store the required resource state in memory until it is needed, we include the running time of preparing such a resource state in this Table as a fair comparison where $p$ is the probability of a successful photon subtraction given some input state.}\label{tab:comparison}
\label{tab:comparison}
\end{table*}
\section{Conclusion}\label{sec:conclusion}
By decomposing the cubic phase gate into a product of non-deterministic operations, each of which only involve a single photon subtraction, we have constructed a  protocol which can approximate the ideal gate using only sequential photon subtractions.  This gate is of vital importance to the field of continuous-variable quantum computing as it enables universal quantum computation when added to the toolbox of the experimentally more simple Gaussian operations.  We have shown that when a photon subtraction is unsuccessful it does not irreversibly destroy the state, and in fact one can continue to attempt this step until it succeeds.  Due to the sequential nature of the photon subtractions, the overall runtime of the protocol scales only as slowly as $\sim 1/p$ where $p$ is the probability of subtracting a photon.

The quality of the proposed gate has been discussed in the presence of realistic detector imperfections which may lead to errors in the implementation of the various $\mathcal U_l$ operations where we have shown that one still obtains a reliable approximation.  Furthermore, we have shown how the approximate cubic phase gate $U_N(\gamma)$ compares to the ideal one when considering how the first two moments of $\hat x,\hat p$ transform for a set of coherent states and for various values of $N$.  We provide an explicit experimental implementation of all components necessary in our protocol and discuss how the probabilistic nature of a photon subtraction comes into play.  

Finally, we compare our implementation of the cubic phase gate to alternative schemes and highlight the similarities as well as the differences in the required resources and nature of the various protocols.  Our protocol offers an experimentally viable method of implementing the cubic phase gate using only a photon detector and gates standard to the alternatives considered in this paper.  Furthermore, we require only sequential photon subtractions as the non-Gaussian element of our protocol.  An open possibility is using this approach to directly implement Hamiltonians of higher power in $\hat x,\hat p$ by modifying the decomposition of $(1+\gamma \hat x^k)$ into a product of $k$ operations.  The largest obstacle in the proposed implementation is likely the requirement that one must protect the state from decoherence and extra noise while attempting to subtract a photon, though advances in quantum memory continue to provide promising results.
\section*{acknowledgements}
K.M. acknowledges support from NSERC.  
\newpage

\end{document}